# Title: PolStar – An Explorer-Class FUV Spectropolarimetry Mission to Map the Environments of Massive Stars

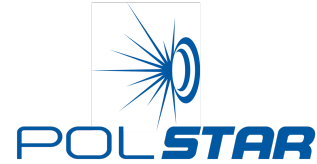

**Type of Activity**: Space Based Project

**Principal Author**:
Name: Paul A. Scowen; Institution: Arizona State University
Email: paul.scowen@asu.edu ; Phone: (480) 965-0938

**Co-authors**: Richard Ignace (ETSU), Coralie Neiner (Observatoire de Paris, Meudon), Gregg Wade (Royal Military College of Canada), Matt Beasley (SwRI), Jon Bjorkman (U. Toledo), Jean-Claude Bouret (LAM, France), Roberto Casini (HAO-NCAR), Tanausu del Pino Alemán (IAC, Spain), Samantha Edgington (LMCO), Ken Gayley (U. Iowa), Ed Guinan (Villanova), Jennifer Hoffman (U. Denver), Ian Howarth (University College London), Tony Hull (UNM), Rafael Manso Sainz (MPS, Germany), Yael Naze (U. Liege), Alison Nordt (LMCO), Stan Owocki (Bartol Inst., U. Delaware), Steve Petrinec (LMCO), Raman Prinja (University College London), Hugues Sana (U. Leuven), Matt Shultz (Bartol Inst., U. Delaware), William Sparks (SETI), Nicole St-Louis (U. Montreal), Clem Tillier (LMCO), Javier Trujillo Bueno (Instituto de Astrofisica de Canarias), Gopal Vasudevan (LMCO), Bob Woodruff (Woodruff Consulting)

## Overview

Massive stars are the most important contributors to galactic cosmic evolution. They provide the raw materials and heavy elements for the interstellar medium, planets and all living matter, and the UV light to illuminate them and catalyze chemistry. Their rapid evolution, extreme luminosity, fundamental dynamical and chemical interactions with their environment will be revealed by *PolStar*.

*PolStar* is an Explorer-class far ultraviolet (FUV) spectropolarimetry mission designed to target massive stars and their environments. *PolStar* will take advantage of resonance lines only available in the FUV to measure for the first time the magnetic and wind environment around massive stars to constrain models of rotation and mass loss that affect the star's evolution and end-of-life state. We will deliver the first 3D tomographic models of the stellar magnetospheres and tie them to known magnetic structures on the stellar surfaces to give us the first complete picture of the stellar environment.

*PolStar* is the first non-solar mission optimized for stellar high resolution UV spectral polarimetry, delivering a spectral resolving power of $R \sim 30,000$, while also measuring all 4 Stokes parameters. *PolStar* is well matched in scope for a timely and scientifically compelling mission. Recent advances in detector technology makes this mission possible now and delivers more than an order of magnitude improvement in sensitivity and spectral resolution over its predecessors.

# 1. Key Science Goals and Objectives

*PolStar* is an Explorer-class far ultraviolet (FUV) spectropolarimetry mission designed to target massive stars and their environments. For unresolved stars, polarimetric data is used to determine system geometry and directly measure stellar magnetism. Many applications fall into the broad categories of scattering polarization or magneto-optical effects. For magnetism, well-known effects include Faraday rotation, synchrotron emission, and the Zeeman and Hanle effects for spectral lines. For scattered light linear polarization from processes such as Rayleigh, dust, or electron scattering tend to be of chief importance.

Many of the outstanding questions surrounding massive star astrophysics find their roots in questions of geometry – finding viewing inclination for binaries to obtain component masses, or for single stars to obtain rotation speeds; determining field axis obliquity for magnetic stars, and field topology (mainly dipole, or more complex); determining opening angles for colliding wind bowshocks; understanding stochastic (e.g., clumps) and organized (e.g., co-rotating interaction regions) in stellar winds; understanding the origin of circumstellar disks; understanding mechanisms of angular momentum transport. These, and more, hinge in part or entirely on discerning system geometry, which is challenging for unresolved sources.

## 1.1 Spectropolarimetric Diagnostics

Spectropolarimetric monitoring involves maximizing the information content available through measurement of a radiation field. For stellar astrophysics this means the measurement of magnetism (strength and topology) and structure (disks, clumps, etc) for unresolved sources.

Polarization data are commonly referenced in terms of a Stokes vector with components I, Q, U, and V (Chandrasekhar 1960). Here I is the total intensity; Q and U are measures of linear polarization, providing both amplitude and position angle (or orientation) information; and V is circular polarization, which for astrophysics is typically associated with sensitivity to magnetism. While I is the measure standard for traditional spectroscopy, relative intensities of $pQ=Q/I$ and $pU=U/I$ are used for linear polarization, and $pV=V/I$ for circular polarization. Crucial for *PolStar* will be the combination of polarization data with monitoring for time-variable measures. The following highlights key diagnostics leveraged through *PolStar's* capability.

### 1.1.1 Continuum Polarization

Continuum polarization from electron scattering allows us to probe source geometry.

One of the best known polarigenic opacities is Thomson scattering for free electrons that produces linearly polarized scattered light. Thomson scattering is the dominant polarigenic (and "gray") opacity in the continuum of massive stars owing to their high temperatures and high levels of ionization. For an unresolved star, a net polarization only results when there is deviation from spherical symmetry. Consequently, linear polarization provides key information about source geometry to complement spectral and temporal data.

More precisely, for linear polarization it requires that the distribution of scattered light not be centro-symmetric. Consequently, electron scattering polarization is typically associated with probing the intrinsic geometry of the source as a deviation from sphericity, such as a non-spherical atmosphere owing to rapid rotation, bipolar jets, circumstellar disks, aspherical mass-loss, stochastic effects such as wind clumping, anisotropic illumination from non-radial pulsations, stellar spots, or even co-rotating interaction regions.

Figures 1a-1c illustrate common applications. Each cartoon figure shows simulated data in the Q-U plane (a standard technique for polarimetry), with the blue dot signifying an offset owing to interstellar polarization. (a) pQ and pU can vary in a linear fashion with time signifying a stationary geometry yet variable in some attribute (e.g., density). (b) Illustrates the case of a binary. The loop pattern signifies changing orientation in the sky; the shape relates to orbital eccentricity and viewing inclination. (c) Random variations, here connected by magenta lines, arise from stochastic variations, such as blob ejections. While pQ and pU are zero in time average, the average total polarization (p=sqrt(pQ^2+pU^2)) does not average to zero. The key insight is that time variable polarimetric behavior provides crucial geometric information to complement spectral data. Note that in all these cases, the ISM polarization is not time variable, meaning variable polarization is always intrinsic to the source.

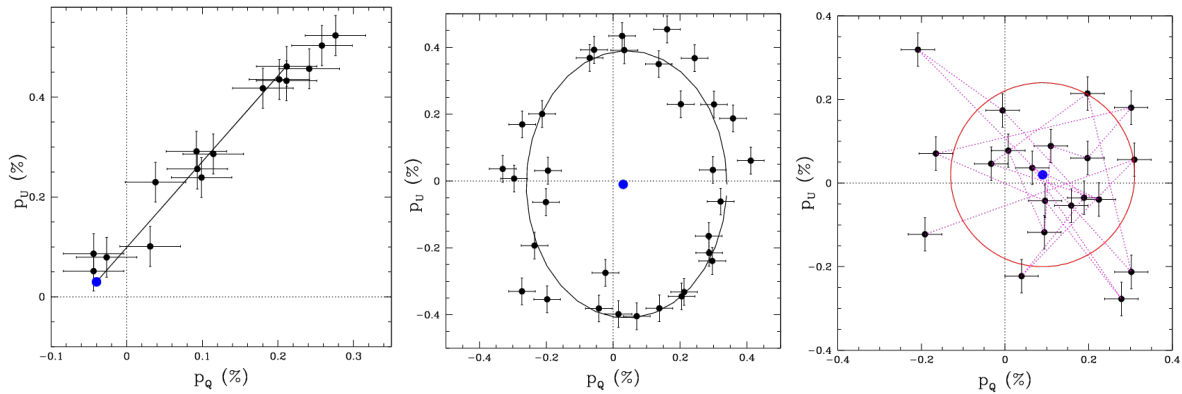

Figure 1: Three illustrative Q-U diagrams for polarimetric variability patterns. Left is a variable source with fixed geometry; middle shows cyclic changes in geometry, such as a binary; right shows stochastic variations, such as wind clumping. (Brown & McLean 1977; Brown et al. 1978; Cassinelli et al. 1987)

1.1.2 Line Polarization

Line polarization permits (a) direct measurement of source magnetism and (b) probing geometry in combination with kinematics within the source.

While Thomson scattering is a continuous opacity, many effects operate in lines. There are two major considerations. First rapid rotation of the star, disk rotation, fast winds, corotating magnetospheres all represent supersonic motions that map the line opacity with different spatial zones at the source via the Doppler shift. This generally leads to changes in the polarization – both in amplitude and position angle – across a resolved spectral line and is used to study the optical depth and geometrical structure of the source. Many of the properties described for continuum linear polarization continue to apply within spectral lines, but now with the addition of kinematical information. Second is a polarigenic response of lines to magnetism. Best know is the longitudinal Zeeman effect, which produces circular polarization that has been applied to the study of stellar magnetism. Less known is the closely associated Hanle effect, which sets in for weaker strengths than the Zeeman effect, leading to alteration of the linear polarization (as opposed to circular) by scattering in the core of a line. The Hanle effect has long been used in the Solar community (Leroy 1977, Landi Degl'Innocenti 1982, Bommier et al. 1994, Casini et al. 2003). While the Zeeman effect has been enormously successful at detecting and detailing surface stellar magnetism, access to UV resonance lines will afford unique opportunities for *PolStar* to probe the circumstellar fields of magnetically channeled and confined winds.

UV spectropolarimetry of hot stars can do far more than simply corroborate the existence of fields of 100 G or more as detected in the optical; it can provide unique information about how the fields extend into the wind, and the nature of the acceleration process at the wind base. Zeeman effect at UV wavelengths produces circular polarization of diminished amplitude relative to the optical. Whereas optical wind lines are in emission, UV lines can be strongly absorbed for massive stars.

In the case of targets with multi-kiloGauss fields, strong magnetic wind confinement, and approximating as a corotating magnetosphere, *PolStar* will be able to detect the Zeeman effect across strong UV lines.

The Hanle effect operates when the Zeeman splitting is small, such that the Larmor frequency associated with the field is of the order 0.1x-10x the A-value of the line. Because it affects polarization by scattering, the Hanle effect is most useful in resonance transitions, which for massive stars are only found in the UV band. For the lines of interest, it offers the ability to diagnose weaker photospheric magnetic fields (1-300 G) than standard Zeeman diagnostic techniques, as well as circumstellar magnetism. The Hanle effect is also much less affected by cancellations from tangled configurations of the magnetic field, as its signal is not destroyed through the polarity reversals that hamper conventional Zeeman effect techniques. Mapping of circumstellar magnetism is achieved with use of multiple resonance lines (thereby sampling spatial regimes of the field tuned to those lines) and with temporal data (thereby sampling different projections of magnetospheres via Doppler shift for rotating and outflowing envelopes). For *PolStar* investigations through the Hanle effect, we plan to adopt diagnostic and modeling tools that have been developed and successfully applied by the solar community.

Crucial to its use as a diagnostic is that the strongest UV resonance lines happen to be Li-like doublets of the type P3/2,1/2–S1/2. The shorter wavelength component ("blue") scatters with a dipole contribution at 50%. The second scatters entirely isotropically. Any polarization observed in the isotropic component must be either interstellar or due to electron scattering at the source. That component acts as a "control" line (e.g., as a monitor of instrumental polarization), which allows to quantify the polarization of the anisotropic component and then be analyzed for the Hanle effect. For *PolStar* science goals, the Hanle effect adds to the diagnostic arsenal in two primary ways. Foremost is measurement of the Hanle effect in photospheric lines for a star other than the Sun for the first time (Manso Sainz & Martinez Gonzalez 2012). Second is the use of the Hanle effect to measure magnetism in the circumstellar environment. Resonance scattering polarization and its modification by the Hanle effect will be used to test models for the interactions between stellar magnetic fields and wind outflow through direct measurements of the circumstellar field, and connect those results to known photospheric fields via the Zeeman effect. The Hanle effect will thus be complementary and ground-breaking in relation to the Zeeman effect for massive stars, something that can only be achieved through UV spectropolarimetry.

## 1.2 Wind/magnetic field interaction and magnetospheric physics

Spectropolarimetric monitoring by *PolStar* will permit mapping of stellar and circumstellar magnetism, with implications for mass-loss physics and angular momentum transport, both of which have ramification for stellar endstates and binary configurations.

In hot stars, fossil magnetic fields with surface strength from 100 G to a few tens of kG, usually dominated by oblique dipoles, have formed and frozen into their radiative envelopes long before the PMS stage (e.g., Alecian et al. 2013). Such fields may contain the imprint of magneto-hydrodynamic processes occurring during early phases of star formation, before the star is visible. Moreover, they have important consequences for later formation phases, especially in their role as intermediaries in accretion and mass-loss processes. During main sequence (MS) and post-main sequence (post-MS) evolution, fossil magnetic fields have been shown to couple strongly to stellar winds, enhancing the shedding of rotational angular momentum through magnetic braking (Townsend et al. 2010). Simultaneously, the presence of the field impedes mass loss, redirecting outflowing wind back toward the stellar surface. As rotation and mass loss are key determinants of the evolution of hot stars, magnetic fields can have an enormous impact (e.g. Petit et al. 2017, Georgy et al. 2018, Kesztheyi et al. 2019).

The results of ground-based spectropolarimetric studies of hot stars at optical wavelengths (e.g., MiMeS, Wade et al. 2016) constitute key observational constraints for the magnetic fossil theory and stellar evolution models. Sophisticated tomographic mapping of stellar surface structures and magnetic fields ispossible; however, due to the current reliance on visible data, such mapping is confined to stellar photospheres. The range of stellar and circumstellar diagnostics provided by *PolStar*, thanks to its UV wavelength coverage, when coupled with the high-cadence, long-term continuous monitoring from space, will allow the extension of indirect mapping methods into the discs and winds of hot stars, providing the first truly 3D views of their magnetospheres and immediate environments (e.g., Simmons 1982, 1983 ; St-Louis et al. 1987 ; Drissen et al. 1987 ; Lupie & Nordsieck 1987 ; Fox 1992; Fox & Henrichs 1994; Li et al. 2009; Ignace et al. 2009).

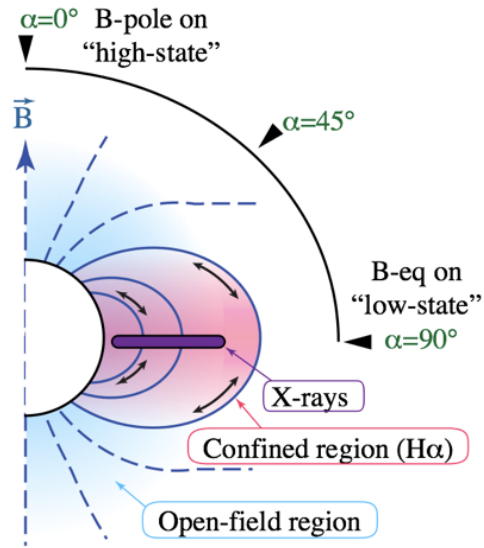

Figure 2: Cartoon illustrating the global structure of the magnetosphere of a slowly-rotating OB star. Adapted from Petit et al. (2013)

By channelling ionised outflows, strong magnetic fields actually control the mass loss of massive stars. Notably, part of the ejected matter falls back onto the star, lowering the net mass loss, but the reduction is known only approximately from 2D and (limited) 3D MHD models of magnetized winds (e.g. ud Doula et al. 2002, 2013). Semi-analytic magnetosphere models (ADM) yield predictions of photometric/polarimetric/resonance line structure and variability that require UV spectroscopy and continuum polarimetry to test. These constraints represent value ingredients for stellar evolutionary models incorporating magnetic fields (e.g. Keszthelyi et al. 2019).

A further consequence is the braking of stellar rotation (e.g. ud Doula et al. 2008), which strongly impacts stellar structure, hence the mixing and stellar chemical yields, and ultimately the stellar evolution itself. Again, this phenomenon has not yet been studied in detail – accurate abundances of key elements like boron, only derivable in the UV, are currently only available for a small number of stars (e.g. Proffitt & Quigley 2001). High-resolution UV spectropolarimetric monitoring will thus not only ensure the full mapping of the magnetosphere, but is also crucial to provide the most sensitive mass-loss and abundance diagnostics.

1.3 Physics, structure, and variability of radiatively-driven winds

UV resonance lines provide a sensitive probe of wind structure via variable P Cygni line absorption. *PolStar's* ability to discern outflow geometry (stochastic or organized) will address imprecision of mass-loss rates among massive stars.

A key defining characteristic of massive stars is their strong winds. High stellar luminosities of energetic UV photons combined with copious metal lines leads to substantial radiative acceleration to drive dense, ionized winds. A massive star can lose 90% of its mass via mass loss during its life (Maeder & Meynet 2000). However, the line-driving process is intrinsically unstable through Doppler shift effects. The result is a runaway scenario that rapidly develops into shock structures permeating the supersonic flow (Owocki & Rybicki 1984), generating spatio-temporal fluctuations in both velocity and density (e.g. Owocki et al. 1988). The UV lines available to *PolStar* encode the properties of this disturbed flow, allowing a focus on two main phenomena: stochastic effects of wind clumping, and large-scale wind structures.

### 1.3.1 Small-scale structures: Clumps and Blobs

Mass loss through radiatively-driven winds strips the hydrogen envelope and bleeds angular momentum from massive stars, drastically altering their evolution and subsequent end-points (e.g. Vink 2012), with direct implications for interstellar feedback and chemical evolution. A critical uncertainty is related to the intrinsically unstable nature of the line-driving mechanism (Owocki et al. 1988; Dessart & Owocki 2003), perhaps amplified by subsurface convection (Cantiello et al. 2009), which leads to strong reverse shocks (Owocki & Puls 1999) that separate fast, low-density wind material from overdense clumps.

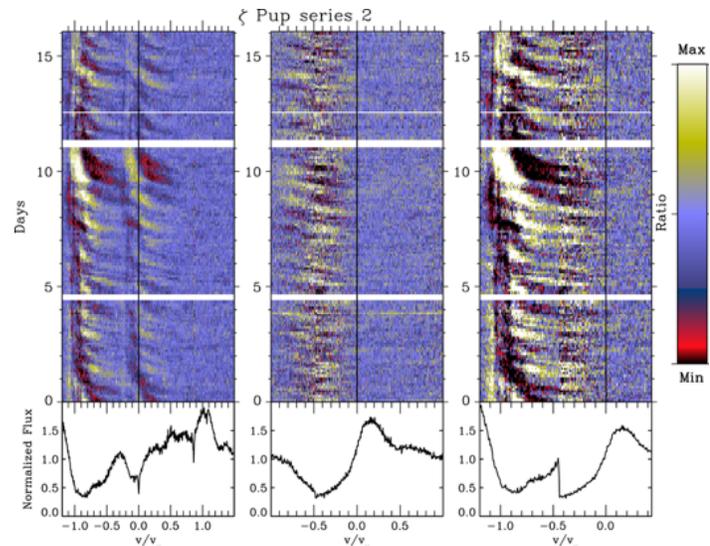

Figure 3: Color representation of the time variability in the UV wind lines of the O supergiant zPup as observed by IUE. Time (in days from the beginning of the series) increases upward. The individual spectra were normalized by a minimum absorption (maximum flux) spectrum so that all changes appear as absorptions. The rest wavelength of the line is shown as a vertical line. The color bar alongside the plot shows the scaling used. Note the cyclical, evolving absorptions features (DACs) that appear first at low velocity and propagate to near the wind terminal speed. Adapted from Massa et al. (1995).

Since many prominent (optical) mass-loss diagnostics are sensitive to the square of the density, they overestimate the mass-loss rate $\dot{M}$ in the presence of wind clumping (Hillier 1991).

Depending on the amount, nature (optically thin versus thick), and spatial/velocity stratification of clumping, a reduction of $\dot{M}$ by a factor of 3 to 10 results, because clumping enhances line emissivity (Bouret et al. 2003, 2005, 2012; Fullerton et al. 2006; Hillier et al. 2003; Sundqvist et al. 2011, 2014, 2018; Sûrlan et al. 2013), including during the late stages of stellar evolution (e.g. Marchenko et al. 2007; Groh et al. 2009). At this level, uncertainty in the absolute mass-loss rates limits our ability to understand the evolution of angular momentum, to predict the initial-to-final mass relation, and the properties of massive stars at death. High signal-to-noise ratio (>100), high cadence (hours) spectroscopic time series of wind-sensitive UV resonance lines for O stars are required to understand the spatio-temporal structure of clumping and its evolutionary dependence.

### 1.3.2 Large-scale structures: Co-rotating Interaction Regions (CIRs)

Aside from stochastic variability, cyclical variability is also commonly observed in massive stars in general and O stars in particular. The UV resonance lines of O supergiants frequently show discrete absorption components (DACs) corresponding to zones of over-absorption throughout the blue-shifted absorption part of the P-Cygni profile (Kaper et al. 1996; Fullerton 2011). They are usually attributed to the presence of large-scale structures present in the stellar wind that lead to optical depth enhancement. Such structures may be due to corotating interacting regions (CIRs) born in the photosphere, typically attributed to surface brightness features, leading to modulations of the launching velocity between adjacent surface regions and ultimately to interaction between slower and faster material. Rotation causes CIRs to develop into a spiral pattern, thereby explaining how DACs are observed to move through P-Cygni profiles with time, and to recur on characteristic timescales.

Several studies have revealed variability in both the wind and photospheric lines of O stars (e.g. Kaufer et al. 2002; Prinja et al. 2006). Whether there is a direct causal link between photospheric (surface) variability and wind variability is still a matter of debate. It is not clear whether the line-driving instability can act on its own, at rather high velocities, or if it is an amplifier of more deeply-seated variations. On theoretical grounds, surface variability is expected if the star experiences non-radial pulsations, as is the case for some O stars (Fullerton et al. 1991; Degroote et al. 2010). Surface inhomogeneities, possibly caused by magnetic spots, could also trigger variability, as was recently proposed by Ramiaramanantsoa et al. (2018) for the O supergiant z Puppis. To test this relationship between photospheric and wind variability further, high-cadence (hours), long-duration (weeks), high high signal-to-noise ratio (>100) UV spectroscopic time series of O stars are required. As many photospheric and wind lines as possible should be monitored simultaneously in order to sample the transition region between the stellar surface and the wind-dominated atmosphere.

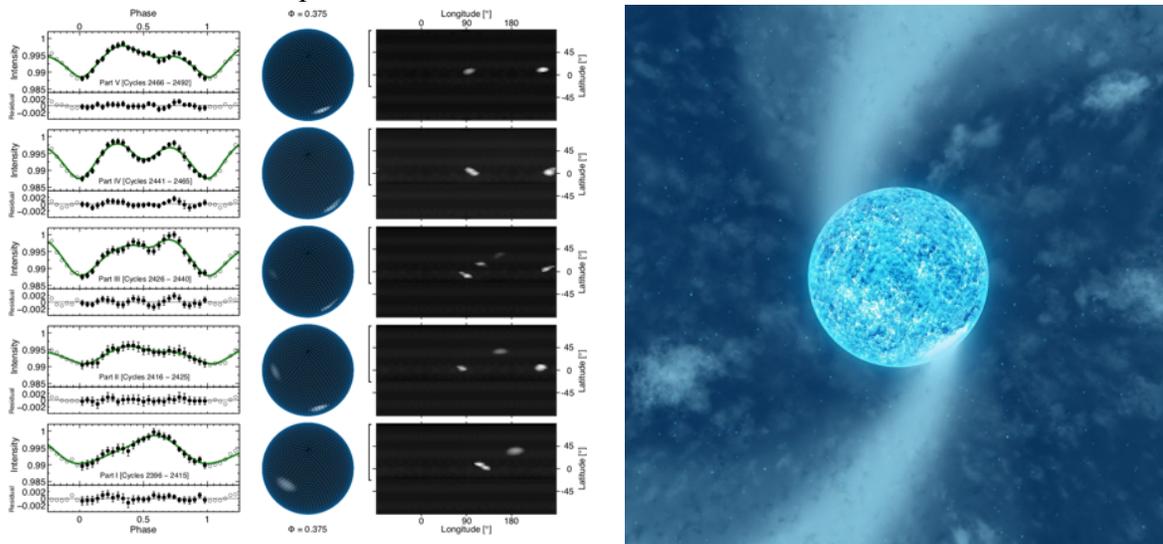

Figure 4: Mapping the photosphere of ζ Pup as observed by BRITE in 2014-2015 using light curve inversion. Time increases upwards. The left panel illustrates the observed light curve (filled circles) during successive parts of the BRITE observing run, along with the reconstructed light curve (green line), with the residuals plotted below the light curves. Then follows a view of the star at rotational phase 0.375 (Middle panel) and the pseudo-Mercator projection of the stellar surface (Right panel). The vertical open brackets on the left of the pseudo-Mercator projections indicate the range of latitudes visible by the observer. The right panel is an artist's conception of the CIRs proposed to be driven by the bright spots inferred from the photometry. From Ramiaramanantsoa et al. (2018).

## 1.4 Massive Binaries

Nearly all (i.e. >70%) massive OB /Wolf-Rayet stars are (or once were) members of binary or multiple star systems. Memberships of stars in binary systems, in particular eclipsing binaries, return crucial information that cannot be known for single stars. The analyses of their orbital radial velocities and light curves provide fundamental physical properties of the component stars. Close interacting massive binary systems with rapidly evolving Wolf-Rayet / OB components serve as "astrophysical laboratories" that yield key estimates of stellar wind densities, mass loss rates, and velocities gleaned from the analysis of colliding winds and mass transfer. When eclipsing, these astrophysical laboratories are used to create detailed maps of the stellar and circumstellar environments.

Massive close binaries are important as progenitors of core-collapse supernovae (SNe Ib, Ic, and II) that enrich, heat, and ionize the ISM. Resulting shock waves and expanding "bubbles" trigger star and planet formation. The end-products of massive star evolution are neutron stars and blackholes. Massive interacting WR+O binary systems are the likely progenitors of gamma ray bursters (GRBs), which requires a rapidly rotating WR star created via binary mass transfer (Woosley & Bloom 2006; Shara et al. 2017). Massive binaries that survive the WR star's explosion are also the progenitors of the compact mergers that create gravitational wave events (Vanbeveren et al. 1997; van den Heuvel 2017). *PolStar* is uniquely equipped to produce a deeper understanding of the mass loss and mass transfer rates of massive binaries.

Polarization effects in close binaries arise from a number of sources and mechanisms: variable polarization from the scattering of stellar flux from the stars' tidally- and rotationally-distorted surfaces; the "reflection effect" in which the light of one star is scattered at the surface of the companion star; and magnetic surface structures as well as magnetic fields created (or embedded) in the outflowing and/or interacting plasmas (e.g., Hoffman et al. 2003; Villar-Sbaffie et al. 2006; Szeifert et al. 2010, Lomax et al. 2012, 2017; Grunhut et al. 2013). The UV spectral region is crucial in these studies because it is very rich in important spectral diagnostics

## 1.5 Origin, Growth, and Evolution of Be Star Disks

Be stars are a class of variable, rapidly rotating, near-main-sequence B stars that have shown (at some time) hydrogen Balmer emission lines, oftentimes double-peaked (see Rivinius et al. 2013 for a review). They are common, with 30% of Galactic B stars being Be stars (Abt 1987; Zorec & Briot 1997), and a frequency that increases both with age (Fabregat & Torrejón 2000; McSwain & Geis 2005) and at low metallicity (about 70% in SMC clusters; Grebel et al. 1992; Keller et al. 1999). In addition to Balmer emission, Be stars have an infrared excess from free-free emission (Gerhz et al. 1974), and they are intrinsically polarized (Coyne 1976). We now know the disks are geometrically thin (Quirrenbach et al. 1997), undergoing Keplerian rotation (Krauss et al. 2012; Wheelwright et al. 2012). Theoretically if stellar material is injected into orbit around the star, viscous forces will automatically spread this material outward forming a so-called viscous decretion disk (Lee et al. 1991) The physical mechanism (turbulent eddy viscosity) that spreads the gas is exactly the same as that operating in pre-main-sequence disks and black hole disks (alpha disk theory of Shakura & Sunyaev 1973), except the material is flowing outward instead of inward (what controls the direction of flow is the radial density gradient; shallow = inflow, steep = outflow). What is unknown is how the material is transferred from the star into the disk, and, in particular, how is enough angular momentum added so that this material attains orbital velocity.

What seems to distinguish Be stars from normal B stars is their rapid rotation. *PolStar* observations and time-dependent monitoring will address key questions for Be stars about injection of material into the disk, how injection is related to stellar pulsations and rapid rotation, how material spreads within the disk, and how disks dissipate. Spectropolarimetry in the line wings of UV resonance lines is sensitive to radial location (via correlation with Keplerian disk velocity with distance). The position angle of the polarization depends on the azimuthal location in the disk, and Q-U loops can be used to map the location of orbiting blobs and to track how they spread and merge into the disk. High S/N UV photospheric line profiles will provide detailed simultaneous observations of the stellar pulsations. For disk dissipation, we can track whether material flows back onto the star as predicted by viscous disk diffusion models (Haubois et al. 2012), or carried away via disk ablation and a disk wind (Kee et al. 2016). We can use *Polstar* to test for the presence of "disk winds" in Be stars. Recently Kee et al. (2016) showed that radiative line-driving can ablate the inner disk, creating a low velocity (~200 km/s) outflow (a disk wind) in a thin layer parallel to the surface of the disk. The disk wind would therefore clear the disk from the inside out and is an alternate mechanism for dissipating the disk and may play an important role in the angular momentum loss from the star.

An important subcategory of Be stars are the gamma Cas analogs. Gamma Cas was the first identified Be star and one of the first extra-solar X-ray sources discovered. However, it soon became clear that its hard and luminous X-ray emission was at strong variance with that observed from the bulk of stars. Galactic X-ray surveys, mainly those carried out by XMM-Newton, led to the discovery of two dozen other Be stars with similar X-ray properties and sharing a narrow range of characteristics (see e.g. Nazé & Motch 2018). The origin of the outstanding X-ray emission of gamma Cas and its analogs remains unknown. Magnetic interaction between the Be star and its decretion disc compete with accretion scenarios onto a white dwarf or neutron star to explain these outstanding features. Both explanations have deep astrophysical implications on massive star evolution and can be tested in the UV with *Polstar*.

## 1.6 Gravity Darkening: Measuring Extreme Rotation with UV Polarization

In the case of rapidly rotating stars, such as Be stars, gravity darkening (von Zeipel 1924) causes the equatorial radiation from the star to be cooler than its polar radiation. If this rapidly rotating star is surrounded by a non-spherical envelope, such as the Be star circumstellar disk, the disk will be the dominant source of the scattered (polarized) flux. The disk is predominantly illuminated by light from the stellar equator, so the scattered flux will have a cooler brightness temperature than the direct (unpolarized) flux from the more poleward regions of the star. Since the polarization is the ratio of the scattered to direct flux, gravity darkening induces an additional color dependence to the linear polarization (Bjorkman & Bjorkman 1994) with the polarization typically falling at UV wavelengths. Bjorkman & Bjorkman show how this effect could be used to determine the rotation rate of the star. Recently, Klement et al. (2015, 2017) produced a detailed model of the Be star Beta CMi in which they report a marginal detection of the gravity darkening effect in optical linear polarization that they used to determine the rotation rate of the star. The detection was marginal primarily due to the lack of current UV linear polarization data, for which *Polstar* will be key in rectifying, and for extending the method to other Be stars.

# 2. Technical Implementation and Design

## 2.1 Instrumentation

The *PolStar* instrument is designed as classic telescope / echelle spectrograph with an inline modulator and analyzer in the entrance of the spectrograph. The instrument is sensitive to light between 1200 Å and 2000 Å. The spectrograph is a cross-dispersed echelle spectrograph that uses a standard 4k x 4k Teledyne e2v CCD 272-84 with has abundant active area for the needs of *PolStar,* and that has been delta-doped to improve its QE across the passband. The echelle provides a resolving power of 30,000 with Nyquist sampled resolution elements. The entire optical structure is athermal with heaters placed to maintain a constant temperature. The *PolStar* instrument is body pointed by the spacecraft ADCS.

## 2.2 Telescope

The telescope is a classic Cassegrain design with a 65 cm aperture at a beam speed of F/13. The optics are made from Zerodur with heritage MgF2/ Al coatings to preserve the reflectivity at 1220 Å. The metering structure is carbon fiber providing an athermal design to simplifying operations with temperature transitions.

## 2.3 Slit, FGS, and Calibration Assembly

The interface between the telescope and the spectrograph is a rotating slit assembly, allowing the selection of two pinholes (0.5 and 2 arcseconds), a dark setting, and a calibration setting, allowing a PtNe lamp to illuminate the detector and provide wavelength calibration in flight. There is a separate lamp in the spectrograph that provides flat fielding for the science detector to monitor long term detector changes from environmental effects.

## 2.4 Polarizer modulator assembly

The polarization modulator assembly consists of two separate components. The first is a set of $MgF_2$ birefringent plates that are rotated to modulate the incoming light beam. The analyzer is a Wollaston prism fabricated out of $MgF_2$ that separates the two output polarizations from the modulator into different angles, allowing a fixed separation in the spectrum. The $MgF_2$ plates are in optical contact, minimizing the number of optical interfaces. By rotating the modulator, the entire stokes vector can be recovered. This design is extremely efficient for measuring the polarization and minimizes optical surfaces with the associated efficiency loss.

## 2.5 Spectrograph

The spectrograph is based on a classic echelle design that operates at the native beam speed of F/13. The light passes through the polarization assembly, then is collimated by an off-axis parabola. The echelle has a groove density of 180 grooves/mm and operates across orders 37 to 61. The light is then cross-dispersed by an off-axis parabola in Wadsworth configuration with a groove density of 750 groove/mm. This places the echellogram across the 4k x 4k detector, using roughly 2000 x 3000 pixels.

# 3. Technology Drivers

The areas of technology used in this design have been deliberately chosen to be as high heritage as possible. We have been working with partners in France at the Observatoire de Paris to focus on a simple spectropolarimeter design with as few moving parts as possible. The current design only has a rotating modulator and a flip mirror for the cal-channel. The two areas of technology that would need further development to achieve TRL 6 by PDR would be:

- The delta doping of the e2v CCDs to achieve the high QE in the FUV necessary for the mission performance
- The optical contact Wollaston prism – to prove that it could survive launch and still be operational in space

Both these development paths are being advanced by either proposed (NASA Heliophysics) or already funded (NASA Astrophysics) Cubesat missions that will advance the TRL of the respective technology.

# 4. Organization, Partnerships and Current Status

The project has been built around the idea of bringing the world's experts on the content areas to the same table. We are working with industrial partner Lockheed Martin out of Palo Alto who have already flown several Explorer class FUV spectropolarimeters for solar work. We are also working with the French group at Observatoire de Paris led by Coralie Neiner, who have done considerable design and development work on next generation polarimeters for use in the FUV and led the design effort for POLLUX on the LUVOIR design concept. Our science team is populated by experts drawn from both the Heliophysics and Stellar Astrophysics worlds in the US, Canada, and Europe to strike an interdisciplinary model that captures techniques from both worlds for the science goals of the mission. We believe this is the best team for this project and have no doubts about our abilities to deliver.

## 5. Schedule

This mission, though requiring a smaller platform, cannot be proposed as a Small Explorer because of the cost of its forefront technology – which caused us to exceed the cost cap by only $15M, or <10%. However we did lay out a detailed schedule to realize the mission and believe that Phases A-D can be executed and completed in 3 years. The mission baseline has a lifetime of 2 years to achieve the science goals described above, but this was the minimum set of goals necessary to prove the approach and yield science deliverables that would open new insight into the environments and interactions with massive stars. A series of extended science mission goals were identified but there is insufficient space here to discuss them.

| WBS # | Name | PI Budget |
|---|---|---|
| Phase A | Phase A | $2,000,000 |
| 1 | Project Management | $4,900,000 |
| 2 | Systems Engineering | $3,500,000 |
| 3 | Safety & Mission Assurance | $1,750,000 |
| 4 | Science (BCD) | $6,000,000 |
| 5 | Science Implementation | $44,671,363 |
| 6 | Spacecraft | $38,494,148 |
| 7 | Mission Ops (Ph E/F All WBS) | $9,500,000 |
| 8 | Launch Services | $50,000,000 |
| 9 | Ground System (B-D) | $3,500,000 |
| 10 | ATLO | $11,235,265 |
| 11 | SC/TD/Other | $0 |
| | **Total of All WBS Elements NO Rsvs.** | $175,550,777 |
| | **Total of Science 4.0** | $6,000,000 |
| | **Total of 1,2,3 (Wraps)** | $10,150,000 |
| | **Total of Development 5,6,9,10** | $97,900,777 |
| | **Total of E/F 7 - No Reserves** | $9,500,000 |
| | **B/C/D Reserves Target** | $34,215,233.10 |
| | **E/F Reserves Target (10%)** | $1,425,000 |
| | **Total PI Cost Plan** | $211,191,010 |
| | **AO PI Cost Cap (A)** | $195,000,000 |
| | **Funds to Employ** | -$16,191,010 |

## 6. Mission Cost Rationale

As mentioned above, the mission design work we had done was targeted at a Small Explorer. The costs for this effort are included below. The costs were based on actually flown hardware, parametric cost estimates, and estimates from reputable vendors to provide elements of the hardware. For a 65cm OTA, the total costs of the heavily descoped mission exceeded the cost cap by at least $15M. However, as a result of this high fidelity analysis, we are confident we can fit this mission (with the descoped science restored) into a Midsized Explorer at the next opportunity. The technologies are relatively well understood and we stand ready to take advantage of the merger of new technology and new scientific knowledge to open a new window on the knowledge of structure and time evolution of stellar environments.

Scowen et al.: **PolStar – An Explorer-Class FUV Spectropolarimetry Mission**Owocki, S. P.; Rybicki, G. B, " Instabilities in line-driven stellar winds. I. Dependence on perturbation wavelength.", 1984ApJ...284..337O

Owocki, Stanley P.; Castor, John I.; Rybicki, George B., « Time-dependent Models of Radiatively Driven Stellar Winds. I. Nonlinear Evolution of Instabilities for a Pure Absorption Model » 1988ApJ...335..914O

Petit, V., et al, 2017, MNRAS, 466, 1052

Prinja, R. K.; Markova, N.; Scuderi, S.; Markov, H.,"The superimposed photospheric and stellar wind variability of the O-type supergiant α Camelopardalis", 2006A&A...457..987P

Proffitt, C, Quigley, M, 2001, ApJ, 548, 429

Quirrenbach, A., et al. 1997, ApJ, 479, 477

Ramiaramanantsoa, Tahina; Moffat, Anthony F. J.; Harmon, Robert; Ignace, Richard; St-Louis, Nicole; Vanbeveren, Dany; Shenar, Tomer; Pablo, Herbert; Richardson, Noel D.; Howarth, Ian D.; Stevens, Ian R.; Piaulet, Caroline; St-Jean, Lucas; Eversberg, Thomas; Pigulski, Andrzej; Popowicz, Adam; Kuschnig, Rainer; Zocłońska, Elżbieta; Buysschaert, Bram; Handler, Gerald; Weiss, Werner W.; Wade, Gregg A.; Rucinski, Slavek M.; Zwintz, Konstanze; Luckas, Paul; Heathcote, Bernard; Cacella, Paulo; Powles, Jonathan; Locke, Malcolm; Bohlsen, Terry; Chené, André-Nicolas; Miszalski, Brent; Waldron, Wayne L.; Kotze, Marissa M.; Kotze, Enrico J.; Böhm, Torsten, "BRITE-Constellation high-precision time-dependent photometry of the early O-type supergiant ζ Puppis unveils the photospheric drivers of its small- and large-scale wind structures", 2018MNRAS.473.5532R

Rivinius, Th., Carciofi, A.C., & Martayan, C. 2013, A&ARv, 21, 69

Shakura, N.I., & Sunyaev, R.A., 1973, A&A, 24, 337

Shara, M., 2017, MNRAS, 464, 2066

Simmons, J., 1982, MNRAS, 200, 91

Simmons, J., 1983, MNRAS, 205, 153

St-Louis, N., et al., 1987, ApJ, 322, 870

Sundqvist, J. O.; Owocki, S. P.; Puls, J., "Atmospheric NLTE models for the spectroscopic analysis of blue stars with winds. IV. Porosity in physical and velocity space", 2018A&A...619A..59S

Sundqvist, J. O.; Puls, J.; Feldmeier, A.; Owocki, S. P., "Mass loss from inhomogeneous hot star winds. II. Constraints from a combined optical/UV study", 2011A&A...528A..64S

Sundqvist, J. O.; Puls, J.; Owocki, S. P., "Mass loss from inhomogeneous hot star winds. III. An effective-opacity formalism for line radiative transfer in accelerating, clumped two-component media, and first results on theory and diagnostics", 2014A&A...568A..59S

Šurlan, B.; Hamann, W. -R.; Aret, A.; Kubát, J.; Oskinova, L. M.; Torres, A. F., "Macroclumping as solution of the discrepancy between Hα and P v mass loss diagnostics for O-type stars", 2013A&A...559A.130S

Szeifert, T., et al., 2010, A&A, 509, L7

Townsend, R., et al, 2010, ApJ, 714, L318

Ud-Doula, A, et al., 2008, MNRAS, 385, 97

Ud-Doula, A, et al., 2013, MNRAS, 428, 2723

Ud-Doula, A, Owocki, S., 2002, ApJ, 576, 413

Van den Heuvel, D., 2017, Handbook of Supernovae (Springer), 1527

Vanbeveren, D., et al. 1997, A&A, 317, 487

Villar-Sbaffie, A., et al., 2006, ApJ, 640, 995
13